\documentclass[aps,prd,10pt,notitlepage,nofootinbib]{revtex4-1}

\usepackage[utf8]{inputenc}
\usepackage{amsmath,amssymb,amsfonts}
\usepackage{ulem}

\usepackage{newpxtext,newpxmath}
\usepackage{graphicx, overpic}
\usepackage{bbold}
\usepackage{bm}
\usepackage{graphicx}
\usepackage[usenames,dvipsnames]{xcolor}
\usepackage{color}
\usepackage[colorlinks=true,linkcolor=blue,urlcolor=magenta,citecolor=magenta]{hyperref}

\usepackage{slashed}
\usepackage[english]{babel}
\usepackage{dcolumn}
\usepackage{pifont}
\usepackage{dsfont,mathrsfs}
\usepackage{cancel}
\usepackage{bigints}
\usepackage{multirow}
\usepackage{natbib}
\usepackage{tikz-feynman}

\usepackage{amsmath, amssymb}
\usepackage{bbold}
\usepackage{makecell}
\usepackage[export]{adjustbox}

\usepackage[bb=boondox]{mathalpha}

\usepackage{orcidlink} 

\newcommand{\fbr}[1]{\left(#1\right)}
\newcommand{\sbr}[1]{\left\{#1\right\}}
\newcommand{\tbr}[1]{\left[#1\right]}
\newcommand{\md}[1]{\left|#1\right|}

\newcommand{\gm}{\gamma}

\newcommand{\n}{\nu}
\newcommand{\munu}{{\mu\nu}}
\newcommand{\alp}{\alpha}
\newcommand{\bt}{\beta}
\newcommand{\ep}{\epsilon}

\newcommand{\sK}{\slashed{K}}

\newcommand{\msd}{\mathscr{D}}
\newcommand{\mcpp}{\mathcal{P}_+}
\newcommand{\mcpm}{\mathcal{P}_-}

\newcommand{\tilkp}{\tilde{K}_+}
\newcommand{\tilkm}{\tilde{K}_-}
\newcommand{\tilskm}{\tilde{\sK}_-}
\newcommand{\tilskp}{\tilde{\sK}_+}

\newcommand{\util}{\tilde{U}}
\newcommand{\gtil}{\tilde{g}}

\newcommand{\wkr}{\omega_k^r}
\newcommand{\wpr}{\omega_p^r}

\newcommand{\MD}{M_D}

\newcommand{\logterm}{\log{\frac{\omega+q}{\omega-q}}}

\newcommand{\Sll}[1]{S_{11}\fbr{#1}}

\definecolor{mygreen}{RGB}{0,150,0}  

\newcommand{\IM}{\text{Im~}}
\newcommand{\RE}{\text{Re~}}

\newcommand{\FB}[1]{(#1)}

\newcommand{\TB}[1]{\left[#1\right]}

\newcommand{\ignore}[1]{}

\allowdisplaybreaks

\begin{document}
	\title{Dynamical color conductivity of a chiral quark-gluon plasma}
	
	\author{Sourav Duari\orcidlink{0009-0006-0795-5186}$^{a,c}$}
	\email{s.duari@vecc.gov.in}
	\email{sduari.vecc@gmail.com}
	
	\author{Nilanjan Chaudhuri\orcidlink{0000-0002-7776-3503}$^{a}$}
	\email{n.chaudhri@vecc.gov.in}
	\email{nilanjan.vecc@gmail.com}

	\author{Pradip Roy\orcidlink{0009-0002-7233-4408}$^{b,c}$}
	\email{pradipk.roy@saha.ac.in}
	
	\author{Sourav Sarkar\orcidlink{0000-0002-2952-3767}$^{a,c}$}
	\email{sourav.vecc@gmail.com}
	
	\affiliation{$^a$Variable Energy Cyclotron Centre, 1/AF Bidhannagar, Kolkata - 700064, India}
	\affiliation{$^b$Saha Institute of Nuclear Physics, 1/AF Bidhannagar, Kolkata - 700064, India}
	\affiliation{$^c$Homi Bhabha National Institute, Training School Complex, Anushaktinagar, Mumbai - 400085, India}

	\begin{abstract}
		
		The dynamical chromoelectric color conductivity of a chiral plasma has been extracted from one loop gluon self energy by using  linear response theory at finite temperature and density. It is shown that due to the $P$ and $CP$ violation the conductivity tensor has an anomalous contribution in addition to the longitudinal and transverse components.  We identify this  anomalous conductivity as chiral chromo-magnetic conductivity. The spectral representations  of real and imaginary parts of the longitudinal and transverse conductivities show marginal variations in chiral plasma as compared to the non-chiral plasma. The static limit of chiral chromo-magnetic conductivity is found to be independent of medium properties reflecting its topological nature. 
		
	\end{abstract}
	
	\maketitle
	\section{Introduction}

	The QCD vacuum comprises of an infinite number of physically equivalent gauge field configurations which are topologically non-trivial bearing  a  distinct winding number~\cite{Shifman:1988zk,Lenz:2001me}. Field configurations known as instantons  mediate transitions between the distinct vacua by tunneling through a potential barrier whose height is of the order of the QCD scale, $\Lambda_{\rm QCD}$, a process known as instanton tunneling~\cite{Belavin:1975fg,tHooft:1976rip,tHooft:1976snw}. However at high enough temperatures such as in the quark-gluon plasma (QGP) created in heavy-ion collisions at RHIC and LHC energies, these transitions can also be induced by another kind of classical thermal excitations known as sphalerons~\cite{Manton:1983nd, Klinkhamer:1984di} which instead of tunneling through the barrier can take the vacuum over the barrier. At these temperatures the sphaleron transition rate can be substantial providing a mechanism for generating $P$- and $CP$-odd bubbles in the QGP. These  gauge field configurations can create or annihilate the chirality of fermions. Thus, if the QGP contains a sufficiently large domain with non-zero winding number one would expect unequal numbers of right- and left-handed quarks and antiquarks leading to a plasma with chiral imbalance~\cite{McLerran:1990de,Moore:2010jd}. This imbalance is usually characterized by a chiral chemical potential \( \mu_5 \) which quantifies the difference in number densities between right- and left-handed quarks.
	
	In the recent past anomaly driven transport phenomena have gained significant attention in research activities ranging from high-energy nuclear physics to condensed matter systems~\cite{Vilenkin:1980fu,Kharzeev:2013ffa,Miransky:2015ava,Kharzeev:2022ydx}. The study of chiral plasma in an electromagnetic field leads to a variety of novel effects such as the Chiral Magnetic Effect (CME)~\cite{Kharzeev:2007jp,Fukushima:2008xe,Kharzeev:2012ph,Kharzeev:2015znc,Koch:2016pzl}, Charge Separation Effect (CSE)~\cite{Son:2004tq,Metlitski:2005pr}, Chiral Electric Separation Effect (CESE)~\cite{Huang:2013iia} and so on.
	Similar mechanisms have also been explored in Dirac and Weyl semimetals~\cite{Son:2012bg,Gorbar:2013dha,Li:2014bha,Cortijo:2016wnf,Kaushik:2018tjj,Sukhachov:2021fkh} where chiral quasiparticles are realized in condensed matter systems.
	
	In QCD plasma, similar to QED plasmas, a finite colour conductivity leads to an induced current as a response of the system to the applied chromoelectric field. Color conductivity has turned out to be an important quantity as it enters directly into the study of long-wavelength (collective) phenomena in the QCD plasma.  A chromo hydrodynamical evolution of the plasma involves transport coefficients such as  color conductivity associated with the color relaxation process. 
	This is demonstrated in Ref.~\cite{Eskola:1992bd} where it is shown that the time evolution of the field energy directly involves the static color conductivity.	
	In Ref.~\cite{Gatoff:1987uf} the role of colour conductivity has also been investigated in nuclear reactions at lower energies. 
	Due to much higher density of partons in the QCD medium created at RHIC and LHC the colour conductivity could be considerably larger and its effects in the hydrodynamic evolution would be more pronounced as it
	enters explicitly into the hydrodynamic evolution equations of the QCD plasma.

	 The static colour conductivity has been addressed by several authors using the semiclassical kinetic theory approach~\cite{Mrowczynski:1989np,Mrowczynski:1988xu,Hwa:1990xg,Heinz:1985qe}. In the relaxation time approximation of transport theory, the static color conductivity $\sigma_c$ can be related to the color relaxation time $\tau_c$ as $\sigma_c=\tau_c\omega_{pl}^2$ where $\omega_{pl}$ is the plasma frequency. Hence $\tau_c$ primarily controls the color conductivity and is closely related to the gluon damping rate~\cite{Selikhov:1993ns}. The static color conductivity and damping rates of quarks and gluons have been estimated at non-zero chemical potential in Ref.~\cite{Hou:1996uq}. The infrared divergence appearing in the color conductivity due to long range magnetic interaction is treated self-consistently by introducing a finite width for quarks and gluons. It is shown that although the damping rates are insensitive  to the chemical potential the color conductivity is enhanced substantially. In Ref.~\cite{MartinezResco:2000pz} the static color conductivity in the leading logarithmic approximation has been computed using the Kubo formalism. It is shown that the results obtained by summing an infinite series of ladder diagrams with the soft collision approximation matches that obtained with QCD kinetic theory. Recently, the calculation of the dynamical color conductivity in a viscous QGP have been performed using QCD kinetic theory~\cite{Jiang:2020lgw}. Numerical estimates of the conductivity show that the color transport process is dominated by the longitudinal part of the conductivity. The inclusion of viscosity influences the real and imaginary parts appreciably in some frequency domains.

In this work we obtain the dynamic chromo-electric conductivity tensor in a chiral QCD plasma from the one-loop gluon polarization function. The latter is calculated using hard thermal loop approximation in the real time formulation of thermal field theory. Through out this article we will use the metric $g^{\mu\nu}=diag(1,-1,-1,-1)$. In Section~\ref{Sec_Formalism}, the formalism for obtaining the colour conductivity is discussed followed by numerical results in Section~\ref{Sec_Results} and discussion in Section~\ref{Sec_Dis}. 
	
	\section{Formalism}\label{Sec_Formalism}
 
	 The induced colour current density $J^a_\mu$ arising from the response of a color plasma to a weak external disturbance within linear response theory is given by~\cite{Kapusta:2006pm,Bellac:2011kqa} 
	 \begin{equation}\label{J}
	 	J_a^\mu(\omega,\vec{q})={_{\rm ret}}\Pi_{ab}^{\mu\nu}(\omega,\vec{q})\ A_\nu^b(\omega,\vec{q})~,
	 \end{equation}
	 where $A^\nu_b$ is the gluon field and $_{\rm ret}\Pi^{ab}_{\mu\nu}$ represents the retarded (color) current-current correlator given by
	 \begin{align}
	 	_{\rm ret}{\Pi}^{\mu\nu}_{ab}(\omega, \vec q)=\int d^{4}X ~e^{i Q\cdot X}~ {_{\rm ret}}{\Pi}^{\mu\nu}_{ab}(X)
	 \end{align}
	 where $Q^\mu=(\omega,\vec q)$, $X^\mu=(t,\vec x)$ and
	 \begin{equation}
	 	{_{\rm ret}}{\Pi}^{\mu\nu}_{ab}(X)=i\theta(t)\langle [ j_a^{\mu}(X), j_b^{\nu}(0) ] \rangle ~.
	 \end{equation}
	   Here $\mu, \nu$ represent Lorentz indices while $a, b$ denote color indices. 	 
	  
	   Again, the colour current density can be expressed through the constitutive relation

	  \begin{equation} 
	J_a^i(\omega,\vec{q}) = \sigma^{ij}_{ab}(\omega,\vec{q}) \ E_b^j ~,
	\label{Jsigma}
\end{equation}
	  where $\sigma^{ij}_{ab}$ is the colour conductivity tensor and $E_b^j$ is the chromoelectric field. In the temporal axial gauge (TAG), defined by $A^0_a=0$, the color electric field can be written as  
	\begin{equation}\label{E}
		E^j_a=i\omega A^j_a .
	\end{equation}
	Substituting this relation into Eq.~\eqref{J}, we obtain 
	\begin{equation}\label{JE_colr}
		J_a^i(\omega,\vec{q})= -\ _{\rm ret}\Pi_{ab}^{ij}(\omega,\vec{q}) \  \dfrac{E^j_b}{i\omega}~.
	\end{equation}
	Now, the retarded current-current correlator is equivalent to the gluon self energy $\Pi^{ab}_{\mu\nu}$~\cite{Kharzeev:2009pj}.
	Since the gluon self-energy is diagonal in the colour space, i.e.  $\Pi^{ab}_{ij}=\delta^{ab}\Pi_{ij}$, Eq.~\eqref{JE_colr} simplifies to
	\begin{equation}\label{JE_colr_less}
		J_a^i(\omega,\vec{q})= -\Pi^{ij}(\omega,\vec{q}) \ \dfrac{E^j_a}{i\omega}~.
	\end{equation}	
	So from now on for brevity of presentation we will suppress the colour index of different quantities such as the induced colour current density, colour conductivity, gluon self-energy \textit{etc}. By comparing the above relation with Eq.~(\ref{Jsigma}), we obtain the colour conductivity in TAG as~\cite{Blaizot:1999fq}
	\begin{equation}\label{conductivity}
		\sigma^{ij}=- \frac{\Pi^{ij}(\omega,\vec{q})}{i\omega}=\frac{i\Pi^{ij}(\omega,\vec{q})}{\omega}
	\end{equation}
\begin{figure}[h]
	
	\begin{tikzpicture}	
		
		\begin{feynman} 
			
			\vertex (a1) {};
			\vertex [right=2cm of a1] (b1);
			\vertex [right=2cm of b1] (c1);
			\vertex [right=2 cm of c1] (d1);
			\vertex [right=1 cm of b1] (p);
			
			\diagram* {
				(a1) -- [gluon, edge label=\(Q\), line width=0.8pt] (b1),
				(c1) -- [gluon, edge label=\(Q\), line width=0.8pt] (d1),
				(c1) -- [fermion, half left, edge label=\(K\), line width=0.8pt] (b1),
				(b1) -- [fermion, half left, edge label=\({P=K+Q}\), line width=0.8pt] (c1),
			};	
			
		\end{feynman}	
		
	\end{tikzpicture}
	\caption{Feynman diagram for one-loop gluon self-energy}\label{one-loop}
\end{figure}
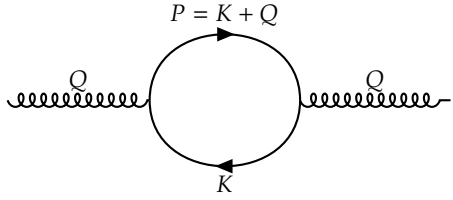


	\begin{figure}

	\end{figure}
	
We now evaluate the one-loop gluon self-energy using the real-time formulation (RTF) of thermal field theory (TFT) where all two-point correlation functions, such as the propagator and self-energy, are represented as $2\times 2$ matrices in thermal space~\cite{Bellac:2011kqa,Mallik:2016anp}. At one loop, the contributions to the gluon self-energy arise from the tadpole diagram involving the four-point gluon vertex, the gluon loop from the three-point vertex, the ghost loop and the fermion loop~\cite{Kapusta:2006pm,Bellac:2011kqa}. 	Among these, the first three contributions remain unaffected by the presence of chiral imbalance whereas the fermion loop acquires non-trivial modifications in such a medium.  In the real-time formulation of TFT that we use, the 11-component of the self-energy matrix is sufficient as shown in~\cite{Duari:2025kxt,Mallik:2016anp}. The 11-component of the gluon polarization tensor shown in Fig.~\ref{one-loop} is given by
	\begin{equation}\label{self energy}
		\Pi_{11}^{\mu\n}(\omega,\vec q)=-ig^2\frac{N_f}{2}\int \frac{d^4K}{(2\pi)^4}\text{Tr}\tbr{\gm^\mu\Sll{P=K+Q}\gm^\n\Sll{K}}
	\end{equation}
	where $g$ is the strong coupling constant, $N_f$ is the number of quark flavours, $Q^\mu=(\omega,\vec q)$ is the external gluon momentum and $K^\mu=(k_0, \ \vec k)$ is the loop momentum. $\Sll{K}$ is the 11-component of the fermionic propagator matrix in chirally imbalanced medium~\cite{Ghosh:2022xbf} which is given by
	\begin{equation}
		\Sll{K}=\msd(k^0+\mu,k) \sum_{r\in \sbr{\pm}}\frac{1}{4kr\mu_5}\tbr{\frac{-1}{\fbr{k^0+\mu}^2-(\wkr)^2+i\ep}-\eta(k^0+\mu) \ 2\pi i\delta\FB{ \ (k^0+\mu)^2-(\wkr)^2 \ }}~.
	\end{equation}
	Here $\eta(x)=\Theta(x)f_+(x)+\Theta(-x)f_-(-x)$, $f_\pm(x)=\TB{\exp\FB{\frac{x\mp\mu}{T}}+1}^{-1}$, $\wpr=p+r\mu_5$, $\wkr=k+r\mu_5$.  $\msd(k^0+\mu,k)$ and $\msd(p^0+\mu,p)$ contain the Dirac structure of the fermion propagators which are given by
	\begin{equation}
		\msd(k^0+\mu,k)= \mcpp \tilkm^2 \tilskp+\mcpm \tilkp^2 \tilskm 
	\end{equation}
	in which $\mathcal{P}_\pm = \frac{1}{2}(1\pm\gm^5)$, $\tilde{K}_\pm^\mu\equiv(k^0+\mu\pm\mu_5,\vec{k})$ and  $\tilde{P}_\pm^\mu\equiv(p^0+\mu\pm\mu_5,\vec{p})$. 
	The general structure of the gluon self-energy in a chirally imbalanced medium is given by~\cite{Nieves:1988qz,Akamatsu:2013pjd,Carignano:2018thu,Carrington:2021bnk,Duari:2025kxt}
	\begin{equation}\label{Eq_Gen_SE}
		\Pi^\munu = \Pi_T R^\munu_T + \Pi_L Q^\munu_L   + \Pi_A P^\munu_A~,
	\end{equation}
	where  $Q^\munu_L$, $R_T^\munu$ and $P_A^\munu$ are orthogonal to each other as well as to $Q_\mu$ and $Q_\nu$ so that gauge invariance is preserved i.e $Q_\mu\Pi^\munu=Q_\nu\Pi^\munu=0$. $Q_L^\munu$, $R_T^\munu$ and $P_A^\munu$ are constructed from the  gluon momentum $q^\mu$, metric tensor $g^\munu$ and thermal bath velocity $U^\mu$. Explicit form of these tensors are as follows,
	\begin{equation}
		Q^\munu_L=\frac{\util^\mu\util^\nu}{\util^2},\hspace{0.5cm}R_T^\munu=\gtil^\munu-Q^\munu_L,\hspace{0.5cm} P_A^\munu=\frac{i}{q}\ep^{\munu\alp\bt}Q_\alp U_\bt~,
	\end{equation}
	where 
	\begin{equation}
		\util^\mu=\gtil^\munu U_\nu,\hspace{0.5cm}\gtil^\munu=g^\munu-\frac{Q^\mu Q^\nu}{Q^2},
	\end{equation}
	and $U^\mu=(1,\vec 0)$ in the local rest frame.
	$\Pi_L$ and $\Pi_T$ are longitudinal and transverse parts of gluon polarization tensor and $\Pi_A$ is the anomalous  contribution which appears due to the chiral asymmetry of the medium. For present purpose only the spatial components of these tensors are required and they are expressed as follows 
	\begin{equation}
		Q^{ij}_L=-\frac{\omega^2}{Q^2}\frac{q^iq^j}{q^2},\hspace{0.5cm}R_T^{ij}=g^{ij}+\frac{q^iq^j}{q^2}=-(\delta^{ij}-\frac{q^iq^j}{q^2}),\hspace{0.5cm} P_A^{ij}=\frac{i}{q}\ep^{ijk}q^k ~.
	\end{equation}
	Using these in Eq.~\eqref{conductivity}, the colour conductivity becomes 
	\begin{equation}\label{conductivity2}
		\sigma^{ij}=\frac{i}{\omega}\tbr{-\Pi_L  \ \frac{\omega^2}{Q^2} \ \frac{q^iq^j}{q^2}       -\Pi_T \ \fbr{\delta^{ij}-\frac{q^iq^j}{q^2}} +\Pi_A \  \frac{i}{q} \ \ep^{ijk}q^k}=\sigma_L \  \hat{q}^i\hat{q}^j+\sigma_T \ \FB{\delta^{ij}-\hat{q}^i\hat{q}^j}-\sigma_{\rm ch} \  \ep^{ijk}\hat{q}^k
	\end{equation}
	where $\sigma_L=-\dfrac{i}{\omega} \ \Pi_L \ \dfrac{\omega^2}{Q^2}$, $\sigma_T=-\dfrac{i}{\omega} \ \Pi_T$ and $\sigma_{\rm ch}=\dfrac{\Pi_A}{\omega}$. The choice of longitudinal and transverse component of the conductivity is made on the basis of gluon momentum. For convenience of the reader we have discussed these in Appendix~\ref{App_A}. Using Eqs.~\eqref{JE_colr_less}, \eqref{conductivity} and \eqref{conductivity2} we can now write the induced colour current density as
	\begin{equation}\label{JE_LTA}
		J^i=\tbr{\sigma_L  \ \hat{q}^i\hat{q}^j+\sigma_T \ (\delta^{ij}-\hat{q}^i\hat{q}^j)-\sigma_{\rm ch} \  \ep^{ijk}\hat{q}^k}E^j~.
	\end{equation}
For physical interpretation, it is convenient to rewrite Eq.~\eqref{JE_LTA} in vector form as
	\begin{equation}
		\vec J = \vec{J}_L+\vec{J}_T+\vec{J}_{ch}=\sigma_L \ \vec{E}_L+\sigma_T \ \vec{E}_T+\sigma_{\rm ch} \ \hat{q}\times \vec{E}=\sigma_L  \  \vec{E}_L+\sigma_T  \ \vec{E}_T+\sigma_{\rm ch}  \ \dfrac{\omega}{q} \ \vec{B}=\sigma_L \  \vec{E}_L+\sigma_T \ \vec{E}_T+ \sigma_{\rm cmc} \ \vec{B} \label{Eq_cond}
	\end{equation}
where $ \sigma_{\rm cmc}=\sigma_{\rm ch} \ \dfrac{\omega}{q}$ and in the second last equality we have used Maxwell's equation (in momentum space)~\cite{Weldon:1982aq}. 	
	The vectorial form of the induced colour current density $\vec J$, as expressed in Eq.~\eqref{Eq_cond}, provides a natural interpretation of the different components of the conductivity. The coefficient $\sigma_L$ corresponds to the longitudinal colour conductivity, while $\sigma_T$ represents the transverse colour conductivity. These two contributions are already present in the absence of chiral imbalance and describe a (colour) current aligned with the chromo-electric field. In contrast, $ \sigma_{\rm cmc}$ characterizes the emergence of a (colour) current along the chromo-magnetic field. This is a novel effect that arises due to the asymmetry between left- and right-handed quarks in the QCD plasma. A similar phenomenon has been identified in a chiral QED plasma, where the associated transport coefficient is known as the chiral magnetic conductivity~\cite{Kharzeev:2009pj,Satow:2014lia}. By analogy, the appearance of $ \sigma_{\rm cmc}$ in the chiral QCD plasma motivates us to identify it as the chiral chromo-magnetic conductivity.
	Now using hard thermal loop (HTL) approximation the expressions for the 	structure functions of the self-energy tensor  are given by~\cite{Akamatsu:2013pjd,Carignano:2018thu,Carrington:2021bnk,Duari:2025kxt}
	\begin{align}
		\Pi_L &= Q_L^\munu \ \Pi_\munu = {\MD^2} \frac{\omega^2-q^2}{q^2}\tbr{1-\frac{\omega}{2q}\logterm} \\
		\Pi_T &= \frac{1}{2}R_T^\munu \ \Pi_\munu= -\frac{1}{2}{\MD^2} \frac{\omega^2}{q^2}\tbr{1+\frac{1}{2}\fbr{\frac{q}{\omega}-\frac{\omega}{q}}\logterm} \\
		\Pi_A &= -\frac{1}{2}P_A^\munu \ \Pi_\munu=-\frac{g^2}{2\pi^2}\frac{N_f}{2} {\mu_5} \frac{\omega^2-q^2}{q}\tbr{1-\frac{\omega}{2q}\logterm}
	\end{align}
	where $\MD$ is the Debye Mass given by
	\begin{equation}
		\MD^2=g^2\frac{T^2}{6}\fbr{N_f+2C_A}+\frac{N_fg^2}{2\pi^2}\fbr{\mu^2+{\mu_5^2}}~.
	\end{equation}

Consequently we arrive at following expressions for conductivities
	\begin{align}
		\sigma_L &= -{i} {\MD^2} \frac{\omega}{q^2}\tbr{1-\frac{\omega}{2q}\logterm} ~\label{Eq_sigL}\\
		\sigma_T &= \frac{i}{2}{\MD^2} \frac{\omega}{q^2}\tbr{1+\frac{1}{2}\fbr{\frac{q}{\omega}-\frac{\omega}{q}}\logterm} \label{Eq_sigT} \\
		 \sigma_{\rm cmc} &= -\frac{g^2}{2\pi^2}\frac{N_f}{2} {\mu_5} \frac{\omega^2-q^2}{ q^2}\tbr{1-\frac{\omega}{2q}\logterm}~.\label{Eq_sigcmc}
	\end{align}
These results are consistent with previous analyses for chiral QED plasma~\cite{Jiang:2020lgw, Satow:2014lia}. 	
Finally, we note that $\sigma_L$, $\sigma_T$, and $ \sigma_{\rm cmc}$ possess both real and imaginary parts which are related through the Kramers–Kronig relations.

\section{Numerical Results}\label{Sec_Results}

In this section, we investigate the frequency dependence of the different conductivities. The motivation for this study is that the behaviour of $\sigma_L$, $\sigma_T$, and $\sigma_{cmc}$ as functions of frequency provides important insights into how the QCD plasma responds dynamically to external chromo-electromagnetic perturbations. In particular, the relative strengths and dispersive properties of the longitudinal, transverse, and chiral chromo-magnetic conductivities reflect the underlying plasma excitations and the role of chiral imbalance. 
For the numerical evaluation, we adopt a representative set of parameters: $N_f=2$, $C_A=1/3$, $\alpha_s=0.3$, $g=\sqrt{4\pi \alpha_s}$, $T=200~\text{MeV}$, $\mu=150~\text{MeV}$ and we take the momentum of the external gluon to be $q=\overline \MD =\MD (\mu_5 = 0)$. These values are chosen to reflect typical conditions relevant for the quark–gluon plasma in the phenomenologically interesting regime.

\subsection{Longitudinal conductivity ($\sigma_L$)}
\begin{figure}[h]
 	\includegraphics[scale=0.32,angle=0]{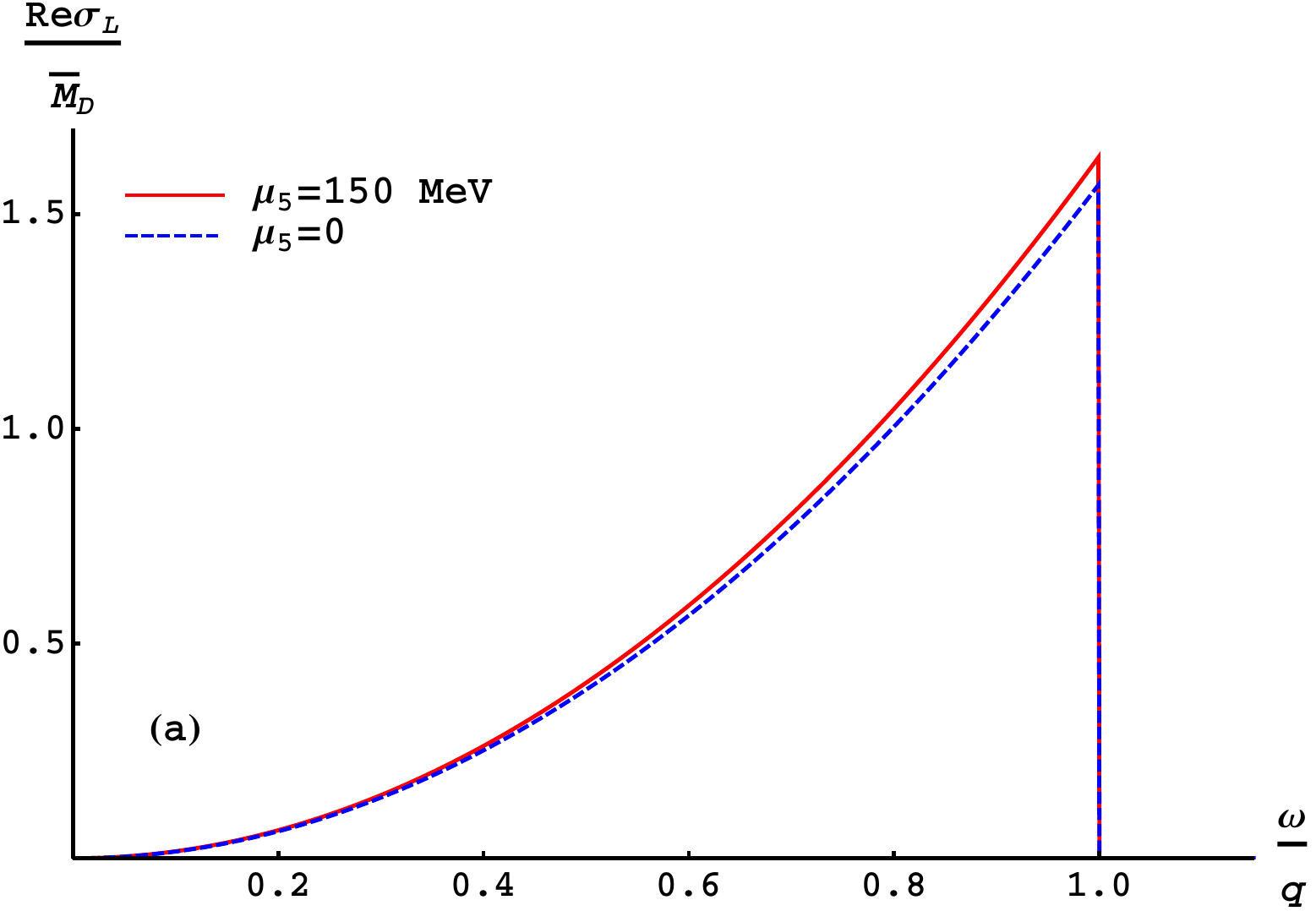} ~~~~  \includegraphics[scale=0.32,angle=0]{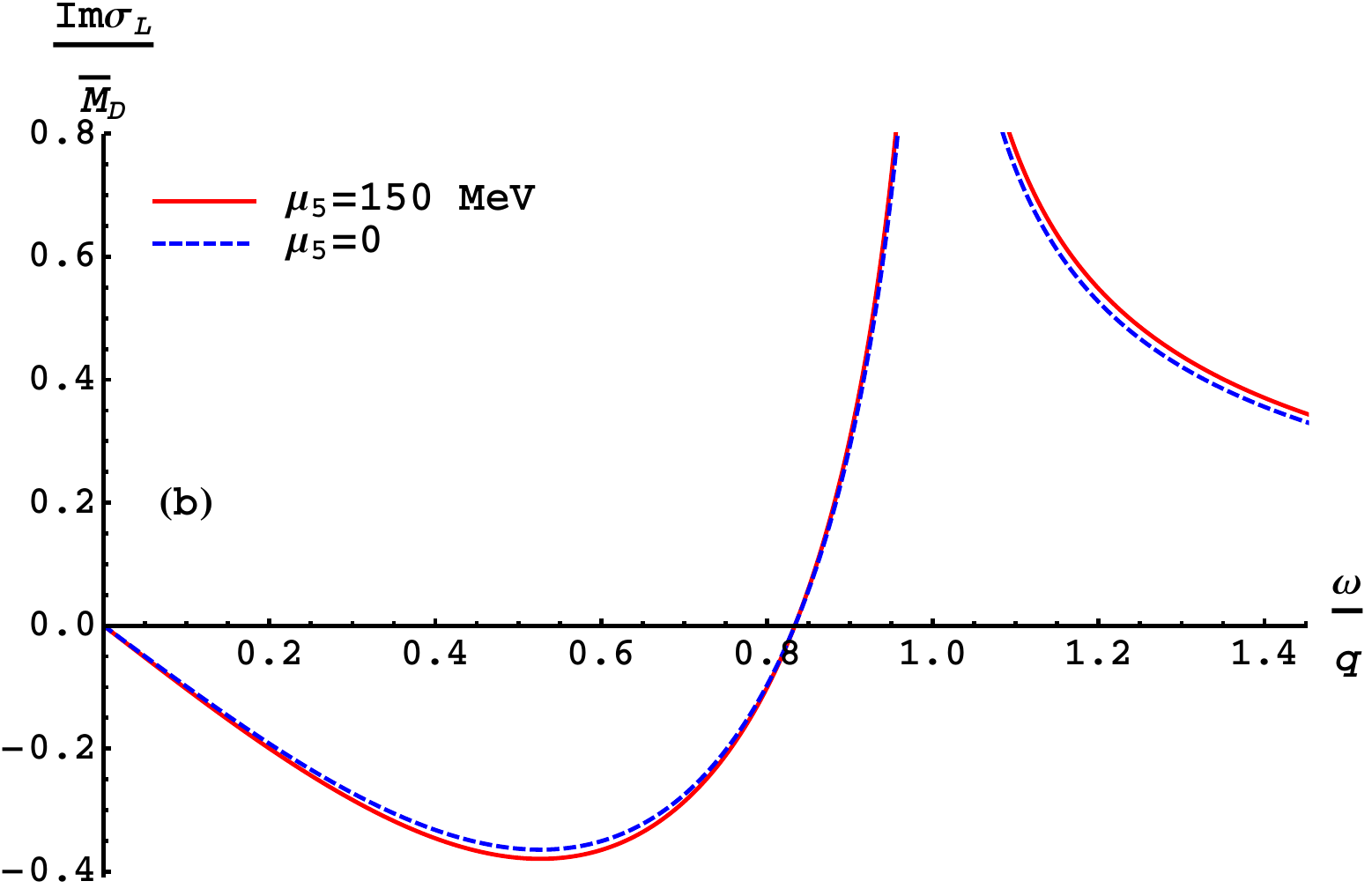}
 	\caption{(a) $\text{Re}\,\sigma_L/\overline{\MD}$  and (b) $\text{Im}\,\sigma_L/\overline{\MD}$  as a function of $\omega/q$ for two values of $\mu_5$ where $\overline{\MD}=\MD (\mu_5 = 0)$. The real part increases monotonically  with frequency in the regime $\omega/q < 1$ in contrast to the imaginary part which decreases initially, increases and diverges due to the logarithmic singularity at $\omega=q$. }
 	\label{Fig_sigL}
 \end{figure}

To investigate the frequency dependence of $\sigma_L$ it is convenient to separate it into real and imaginary components as these parts carry distinct physical implications that will be discussed shortly. The logarithmic structure appearing in the HTL expressions for the self-energy can be analytically continued using the standard identity  
 \begin{equation}
 	\log\frac{\omega+q+i\epsilon}{\omega-q+i\epsilon} 
 	= \log\md{\frac{\omega+q}{\omega-q}} - i\pi \ \Theta(q^2-\omega^2)~,\label{Eq_log}
 \end{equation}  
which separates the real and imaginary contributions explicitly. Employing this relation, the longitudinal conductivity can be written as
 \begin{align}
 	\text{Re}\,\sigma_L &= \dfrac{\MD^2}{2q} \ \dfrac{\omega^2}{q^2} \ \pi \ \Theta(q^2-\omega^2),\label{Eq_ResigL} \\
 	\text{Im}\,\sigma_L &= -\frac{\MD^2}{q} \  \frac{\omega}{q}\tbr{1-\frac{\omega}{2q}\log\md{\frac{\omega+q}{\omega-q}}}~.\label{Eq_ImsigL}
 \end{align}
 The presence of the step function $\Theta(q^2-\omega^2)$ ensures that the real part of the longitudinal conductivity is non-vanishing only for space-like momenta ($\omega < q$), while it vanishes identically for time-like kinematics ($\omega > q$).  Fig.~\ref{Fig_sigL}~(a) illustrates the behaviour of the real part of $\sigma_L$ as a function of frequency scaled by $\overline{\MD}$. For both zero and finite chiral chemical potential $\mu_5$ the real part increases monotonically with frequency in the regime $\omega/q < 1$, reflecting its parabolic dependence on $\omega$. Once $\omega/q$ exceeds unity the real part disappears due to the step-function constraint. Importantly, the effect of a finite chiral chemical potential is to enhance the overall magnitude of the longitudinal conductivity. Moreover, this enhancement becomes more pronounced as $\omega$ increases. The underlying reason is that a nonzero $\mu_5$ contributes positively to the Debye mass thereby strengthening  screening of the medium and consequently amplifying the conductivity. Quantitatively the vector and chiral chemical potentials play a similar role since they appear additively in the expression for $M_D^2$. At $T=200$ MeV an increase of any one from 0 to 150 MeV at vanishing value of the other, results in about $4\%$ increase in the value of $M_D^2$.
The behaviour of the imaginary part of $\sigma_L$ as a function of $\omega/q$ is presented in Fig.~\ref{Fig_sigL}~(b). Unlike the real part, which vanishes outside the space-like domain, the imaginary component remains nonzero for all values of $\omega$ and exhibits a distinctly non-monotonic dependence on frequency. This feature can be understood by considering the small-$\omega$ expansion of Eq.~\eqref{Eq_ImsigL}, which yields
 \begin{equation}
 	\IM \sigma_L \approx \dfrac{\MD^2}{q} \tbr{ -\dfrac{\omega}{q} + \fbr{\dfrac{\omega}{q}}^3+ \dfrac{1}{3} \fbr{\dfrac{\omega}{q}}^5 + \cdots}~.
 \end{equation}
 The series clearly shows that $\IM\sigma_L$ initially decreases with increasing frequency before higher-order terms become relevant leading to a change in behaviour. On the other hand, at $\omega = q$ the imaginary part diverges because of the logarithmic singularity in Eq.~\eqref{Eq_ImsigL}. Consequently, $\IM\sigma_L$ must attain a minimum within the interval $0<\omega/q<1$ which is indeed visible in Fig.~\ref{Fig_sigL}~(b). A direct analysis of Eq.~\eqref{Eq_ImsigL} confirms that this minimum occurs at $\omega/q \approx 0.53$. The difference between the zero and non-zero $\mu_5$ results is however marginal because of reasons stated above. For $\omega \gg q$ one can write
 \begin{equation}
\IM \sigma_L \approx {\MD^2}\tbr{ \dfrac{1}{\omega} + \dfrac{q^2}{5 \omega^3} + \cdots}~.
 \end{equation}
 This explains the fact that \(\IM \sigma_L\) falls off as \(1/\omega\) for large values of $\omega$ as observed in Fig.~\ref{Fig_sigL}~(b).

\subsection{Transverse conductivity $\sigma_T$}
Using Eq.~\eqref{Eq_log} in Eq.~\eqref{Eq_sigT}, the transverse conductivity can be decomposed into real and imaginary parts as
\begin{align}
	\text{Re}\,\sigma_T &=\frac{\MD^2}{4q}\fbr{1-\frac{\omega^2}{q^2}}\pi \, \Theta(q^2-\omega^2),\label{Eq_ResigT} \\
	\text{Im}\,\sigma_T &= \frac{\MD^2}{2q} \frac{\omega}{q}\tbr{1+\frac{1}{2}\fbr{\frac{q}{\omega}-\frac{\omega}{q}}\log\md{\frac{\omega+q}{\omega-q}}}~.\label{Eq_ImsigT}
\end{align}
\begin{figure}[h]
	\includegraphics[scale=0.32,angle=0]{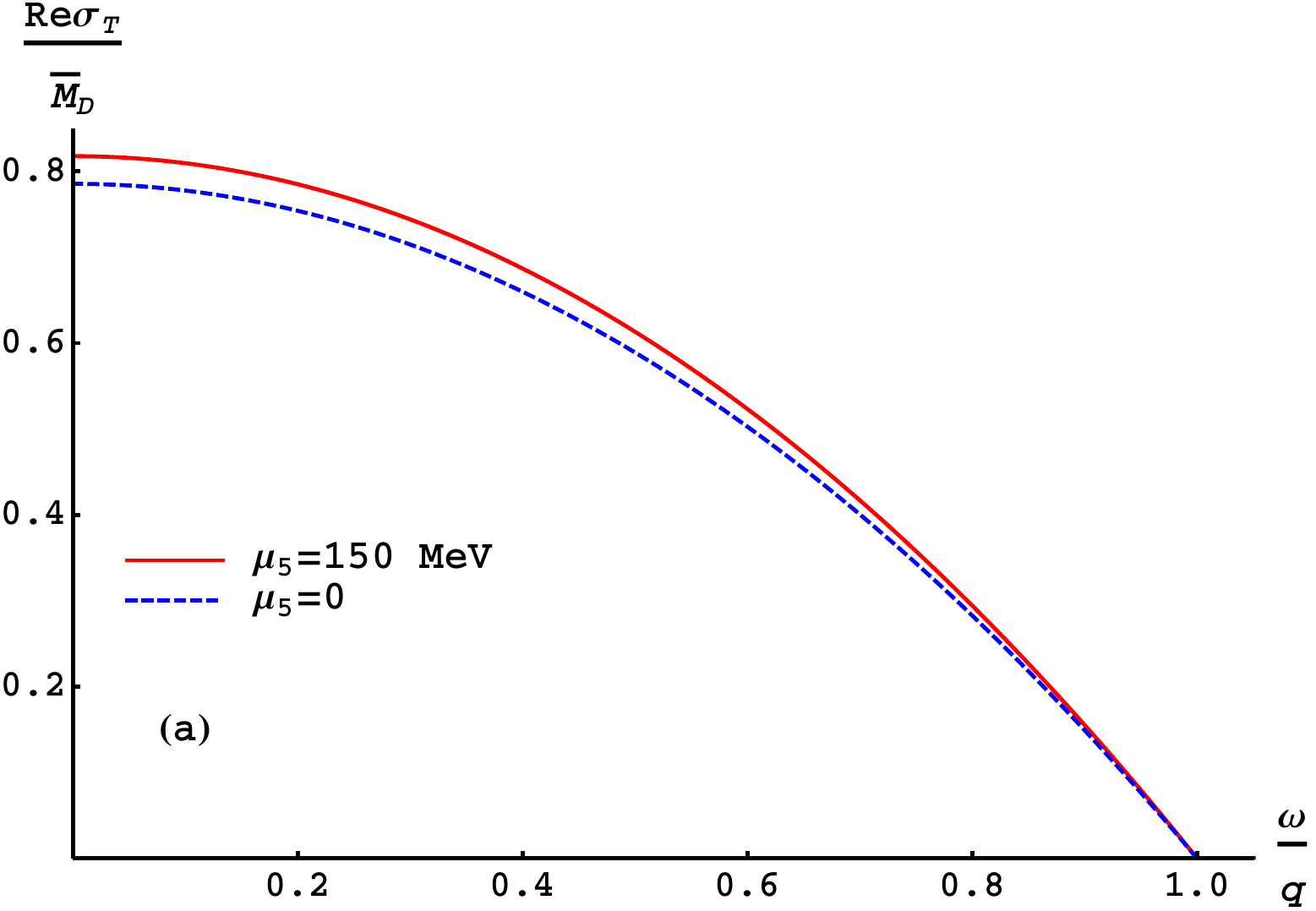} ~~~~ 	\includegraphics[scale=0.32,angle=0]{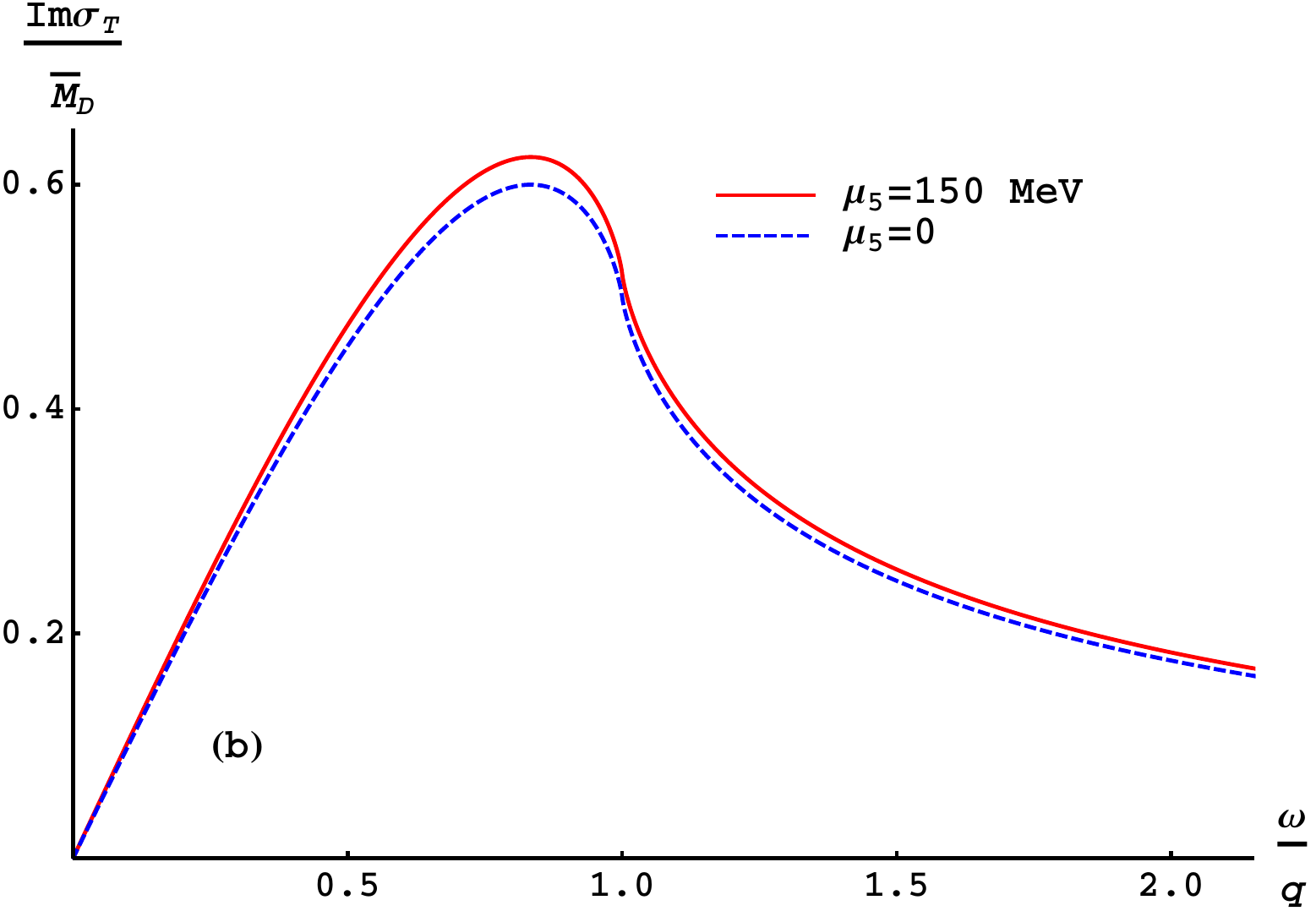}
	\caption{(a) $\text{Re}\,\sigma_T/\overline{\MD}$  and (b) $\text{Im}\,\sigma_T/\overline{\MD}$  as a function of $\omega/q$ for two values of $\mu_5$ where $\overline{\MD}=\MD (\mu_5 = 0)$. The real part decreases monotonically  with frequency and vanishes at $\omega/q = 1$. The imaginary part increases with frequency, reaches a maximum, and then decreases at larger values of $\omega/q$.}
	\label{Fig_ResigT}
\end{figure}
The real part of the transverse conductivity is non-vanishing only for space-like momenta as enforced by the step function in Eq.~\eqref{Eq_ResigT}. As shown in Fig.~\ref{Fig_ResigT}~(a), $\text{Re}\,\sigma_T$ decreases monotonically with increasing frequency and vanishes at $\omega=q$. The overall magnitude is enhanced when a finite chiral chemical potential is present, reflecting the increase of the Debye mass $\MD$ with $\mu_5$. However, this enhancement becomes progressively smaller as the frequency grows, and the two curves converge near $\omega=q$.
The behaviour of the imaginary part, plotted in Fig.~\ref{Fig_ResigT}~(b), is qualitatively different from that of the real part. Specifically, $\text{Im}\,\sigma_T$ increases with frequency, reaches a maximum, and then decreases at larger values of $\omega/q$. A direct analysis of Eq.~\eqref{Eq_ImsigT} confirms that this maximum occurs at $\omega/q \approx 0.83$. This non-monotonic dependence can be understood by expanding Eq.~\eqref{Eq_ImsigT} at small $\omega$, which gives
\begin{equation}
	\IM \sigma_T \approx \dfrac{\MD^2}{q} \tbr{ \dfrac{\omega}{q} - \dfrac{1}{3}\fbr{\dfrac{\omega}{q}}^3 -  \dfrac{1}{15} \fbr{\dfrac{\omega}{q}}^5 + \cdots}~.
\end{equation}
The leading term shows a linear rise at low frequency, while higher-order corrections eventually lead to a turnover, explaining the observed maximum. On the other hand for $\omega\gg q$ one one can write down the expansion of $\IM \sigma_T$ as
\begin{equation}
	\IM \sigma_T \approx \dfrac{\MD^2}{3} \tbr{ \dfrac{1}{ \omega}  + \dfrac{q^2}{5\omega^3} +\cdots }  
\end{equation}
 Hence \(\IM \sigma_T\) falls off as \(1/\omega\) at large frequency as evident from Fig.~\ref{Fig_ResigT}~(b). As with the real part, a finite chiral chemical potential enhances the overall magnitude of $\text{Im}\,\sigma_T$, but does not alter its qualitative structure.
 
It is easy to check that the real and imaginary parts given by Eqs.~\eqref{Eq_ResigL} and \eqref{Eq_ImsigL} respectively for the longitudinal conductivity and  by  Eqs.~\eqref{Eq_ResigT} and \eqref{Eq_ImsigT} for the transverse conductivity are related to each other through the Kramers-Kronig relations.

\subsection{Chiral magnetic conductivity $ \sigma_{\rm cmc}$}

\begin{figure}[h]
\includegraphics[scale=0.45,angle=0]{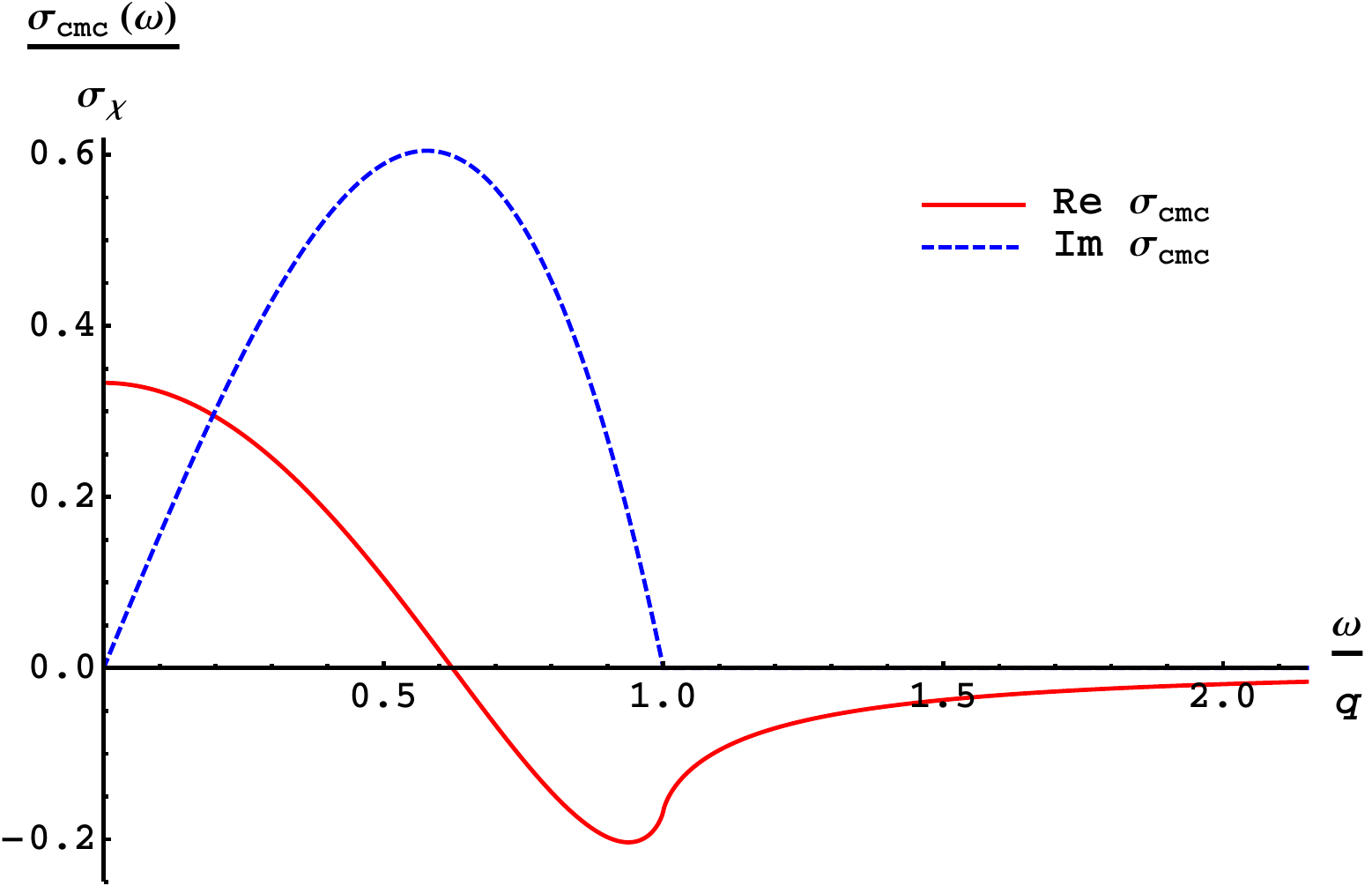} 
\caption{Frequency dependence of real and imaginary part of the  chiral chromo-magnetic
	conductivity $\sigma_{\rm cmc}$ scaled by $\sigma_\chi$, its value in the static limit, at $q=\mu_5$.  Real part at $\omega \ne 0$ does not receive any contribution from the imaginary part at $\omega = 0$. As a result, for any temperature the real part of the chiral magnetic conductivity drops from $\sigma_\chi$ at $\omega = 0$ to $\sigma_\chi/3$ just away from $\omega = 0$. $\IM \sigma_{\rm cmc}$ is well behaved at large $\omega$ and is nonvanishing only for space-like kinematics ($\omega<q$).}
	\label{Fig_rd}	
\end{figure}
From Eq.~\eqref{Eq_sigcmc} the real and imaginary parts of the chiral chromo-magnetic conductivity can be expressed as  
\begin{align}
	\text{Re}\, \sigma_{\rm cmc} &= -\frac{g^2}{2\pi^2}\frac{N_f}{2} \,\mu_5 \,
	\frac{\omega^2-q^2}{q^2}\left[1-\frac{\omega}{2q}\log\Bigg|\frac{\omega+q}{\omega-q}\Bigg|\right], \label{Eq_RsigX}\\
	\text{Im}\, \sigma_{\rm cmc} &= -\frac{g^2}{2\pi^2}\frac{N_f}{2} \,\frac{\mu_5}{2}\,
	\frac{\omega}{q}\,\frac{\omega^2-q^2}{q^2}\,\pi \,\Theta(q^2-\omega^2).\label{Eq_IsigX}
\end{align}
The above decomposition shows that $\sigma_{\rm cmc}$, unlike the longitudinal and transverse conductivities, is entirely proportional to the chiral chemical potential $\mu_5$. This feature is a manifestation of the fact that chiral chromo-magnetic conductivity arises solely due to the imbalance between left- and right-handed quarks. 
Concentrating on the real part, one can see that in the static limit $\omega \to 0$, we get 
\begin{equation}
	 \sigma_\chi :=\lim_{\omega\to 0} \, \text{Re}\, \sigma_{\rm cmc} \;=\; \frac{g^2}{2\pi^2}\frac{N_f}{2}\,\mu_5 ,
\end{equation}
which is independent of temperature $T$ and vector chemical potential $\mu$.
This reflects the topological nature of the chiral chromo-magnetic conductivity as the coefficient is fixed by the axial anomaly and remains unaffected by loop corrections from gluons or photons~\cite{Fukushima:2008xe,Kharzeev:2009pj,Satow:2014lia,Landsteiner:2013aba}.
However, in contrast to $\sigma_L$ and $\sigma_T$, the real part of $\sigma_{\rm cmc}$ does not vanish at large values of $\omega$. Instead it saturates to some finite value. Thus the $\RE \sigma_{\rm cmc}$ and $\IM \sigma_{\rm cmc}$  given by Eqs.~\eqref{Eq_RsigX} and \eqref{Eq_IsigX} respectively are not related by  Kramers-Kronig formula. $\IM \sigma_{\rm cmc}$ is well behaved at large $\omega$. It is nonvanishing only for space-like kinematics ($\omega<q$) which is obvious from the $\Theta$-function appearing in Eq.~\eqref{Eq_IsigX}. So in Fig.~\ref{Fig_rd} we have plotted $\RE \sigma_{\rm cmc}$ as obtained by employing the Kramers-Kronig relation. From this relation it follows that  the real part at $\omega \ne 0$ does not receive any contribution from the imaginary part at $\omega = 0$. As a result, for any temperature the real part of the chiral magnetic conductivity drops from $\sigma_\chi$ at $\omega = 0$ to $\sigma_\chi/3$ just away from $\omega = 0$ as evident from Fig.~\ref{Fig_rd} for the $\RE \sigma_{\rm cmc}$. This is consistent with the observations made in~\cite{Kharzeev:2009pj,Landsteiner:2013aba} for the case of Abelian chiral plasma.

\section{Discussions}\label{Sec_Dis}

In this work we have studied the dynamical color electric conductivity of a quark gluon plasma with chiral imbalance. Within the ambit of hard thermal loop approximation in the real time thermal field theory one loop gluon self energy is calculated using diagrammatic approach. The conductivity is then obtained using the constitutive relation and linear response theory. Due to $P$ and $CP$ violation an anomalous contribution to the conductivity is obtained in addition to the usual longitudinal and transverse contributions. We identify the anomalous part as the chiral chromo-magnetic conductivity. We show the dynamical behaviour of various parts of the color conductivity. It is found that the effect of the chiral chemical potential is marginal in both the real and imaginary parts of the transverse and longitudinal conductivities except in a very narrow region of frequency ($\omega$). We have obtained the leading order chiral chromo-magnetic conductivity. It has been explicitly shown that in the static limit the chiral chromo-magnetic conductivity is independent of temperature and vector chemical potential reflecting its topological nature. Using Kramers-Kronig formula the frequency dependence of both real and imaginary parts of $\sigma_{\rm cmc}$ is shown. All the results are compatible with the previous observations made for Abelian chiral plasma in Refs.~\cite{Kharzeev:2009pj,Landsteiner:2013aba,Satow:2014lia}.

Few comments on gauge invariance are in order here. In this work we have evaluated the conductivity tensor to lowest order in the temporal axial gauge using the HTL method. The gauge invariance of observables obtained using HTL techniques to leading order have been shown by Gagnon and Jeon for the case of electrical conductivity~\cite{Gagnon:2006hi} as well as for shear viscosity~\cite{Gagnon:2007qt}. Again, Braaten and Pisarski~\cite{Braaten:1989mz} has shown that to leading order in $g$ the quark and gluon damping rates are gauge invariant. (See also~\cite{Kobes:1990dc} for a general discussion on gauge independence of observables within HTL.)
	 
We conclude by pointing out that the chiral chromomagnetic conductivity discussed here is a non-Abelian generalization of the well known chiral (electro)magnetic effect, both of which are a consequence of the non-trivial topological vacuum structure of QCD. In the lowest order calculation shown here the QCD coupling and flavor factors replace the QED coupling in the anomalous part whereas the gluon and ghost loop contributions modify the prefactor appearing in the expression for the Debye screening mass in case of transverse and longitudinal components of the chromoelectric conductivity tensor. The  Debye mass in this case turns out to be an order of magnitude larger than the Abelian case for parameters used in this work, resulting in a much shorter screening length for the chromoelectric fields. It is worth mentioning that the calculation of transport coefficients in non-Abelian plasmas beyond leading order may be quite challenging ~\cite{Arnold:1998cy,Arnold:1999uy,Gagnon:2007qt,Gagnon:2006hi}. Note that in general, non-Abelian plasmas are found to be poor color conductors due to over-damping of the collective color modes~\cite{Selikhov:1993ns,Heiselberg:1994px}.
 
\appendix

\section{Spatial part for the longitudinal and transverse components with respect to the gluon momentum } \label{App_A}

Here we define the longitudinal and transverse components with respect to the gluon momentum $\vec{q}$. The longitudinal direction is $\hat{q}=\frac{\vec{q}}{q}$. So any vector $\vec{v}$ has a longitudinal component given by $v_L=\frac{\vec{q}\cdot\vec{v}}{q}=\hat{q}\cdot\vec{v}$. Now its component transverse to the momentum $\vec q$ is given by $\vec{v}$ is $v_T^i=v^i-v_L \hat{q}^i=(\delta^{ij}-\hat{q}^i\hat{q}^j) \ v^j$. 
So $\vec{v}$ can be written as
\begin{equation}
	v^i=\frac{\vec{q}\cdot\vec{v}}{q} \ \hat{q^i}+(\delta^{ij}-\hat{q}^i\hat{q}^j) \ v^j~.	
\end{equation} 
Similarly we can define any symmetric tensor $S^{ij}$ as 
\begin{equation}
	S^{ij}=S_L \hat{q}^i\hat{q}^j+S_T(\delta^{ij}-\hat{q}^i\hat{q}^j)
\end{equation}

	
	\bibliography{main} 
	
\end{document}